\documentclass[notitlepage,10pt]{article}
\usepackage{latexsym}
\newcommand{\nlc}{$e^{+}e^{-}$ linear collider}
\newcommand{\etal}{\textit{et al.}}
\def\pl#1#2#3{\frenchspacing{\it Phys. Lett. }{\bf #1}, #2 (19#3)}
        
\def\prl#1#2#3{\frenchspacing{\it Phys. Rev. Lett. }{\bf #1}, #2 (19#3)}

\def\pr#1#2#3{\frenchspacing{\it Phys. Rev. D}{\bf #1}, #2 (19#3)}

\def\np#1#2#3{\frenchspacing{\it Nucl. Phys. }{\bf #1}, #2 (19#3)}

\def\zp#1#2#3{\frenchspacing{\it Z.~Phys. C}{\bf#1}, #2 (19#3)}

\def\ib#1#2#3{\frenchspacing{\it ibid. }{\bf #1}, #2 (19#3)}

\def\sciam#1#2#3#4{\frenchspacing{\it Sci. Am. }{\bf #1}, (\ifcase#3\or January\or February\or March\or April\or May\or
June\or July\or August\or September\or October\or November\or 
December\fi, 19#4), p.~#2}

\def\phystoday#1#2#3#4{\frenchspacing{\it Phys. Today }{\bf #1}, #2 (\ifcase#3\or January\or 
         February\or March\or April\or May\or June\or July\or August\or 
         September\or October\or November\or December\fi, 19#4)}
\newcommand{\hepph}[1]{hep--ph/#1}
        
\relax
\begin{document}
\begin{flushright}
	FERMILAB--CONF--96/215--T
\end{flushright}
\vspace*{24pt}
	\begin{center}
		{\LARGE Top Priorities:} \\ {\large Questions for Snowmass '96}\\[12pt]
	
		{\large Chris Quigg }\\ Fermi National Accelerator Laboratory \\
		P.O. Box 500, Batavia, Illinois 60510	\\[12pt]
	
	May 28, 1996
	\vspace*{24pt}
	\end{center}

\begin{enumerate}
	\item  Update the analysis of constraints on the Higgs-boson mass 
	that result when precision electroweak measurements at the $Z^{0}$ 
	pole are supplemented by precision measurements of the top-quark 
	mass and the $W$-boson mass.  For given values of $\delta M_{W}$, 
	exhibit the sensitivity of expectations for $M_{H}$ to various 
	assumed values of $\delta m_{t}$.

	\item  Make a critical examination of prospects for $\delta M_{W}$ 
	at the Tevatron, LEPII, and the LHC, and for $\delta m_{t}$ at the 
	Tevatron, the LHC, and an \nlc.  Pay special attention to what could 
	go wrong, including implicit (physics) assumptions that might not be
	fulfilled.  What are the ultimate theoretical limitations to these 
	measurements?

	\item  Understand (resolve) the differences among the competing 
	calculations of the cross section for $\bar{p}p \rightarrow 
	t\bar{t}+\hbox{ anything}$ in QCD.  How well will the cross section 
	be measured at the Tevatron and the LHC?

	\item  How secure is the conclusion (from CDF analysis) that the 
	decay $t \rightarrow b+W$ accounts for $87^{+13+13}_{-30-11}\%$ of 
	top decays?  How much can the measurement be improved?

	\item  How well can the Tevatron, the LHC, and an \nlc\ measure 
	$|V_{tb}|$ if top is normal?  How well can we establish that top is 
	normal (i.e., has no anomalous couplings)?
	
	\item How can spin correlations aid the search for new physics, 
	including anomalous couplings and CP violation?

	\item  Find a strategy to place an upper bound on the top lifetime by 
	direct observation.
	
	\item Develop strategies for determining the total width of top, 
	$\Gamma(t\rightarrow \hbox{ all})$, at the Tevatron, LHC, and a 
	near-threshold \nlc.

	\item  Quantify expectations for the ``dead cone'' for energetic top 
	quarks.  Develop a strategy for investigating the dead cone for $b$ 
	and $t$ quarks.

	\item  How many of the conjectures about new states with masses 
	below $m_{t}$, or even below $M_{W}$, can be ruled out (or 
	vindicated) now?  What will it take to do better in Run II?  What are 
	the implications for the TeV33 program?
	
	\item How well can $t\bar{t}$ invariant mass distributions be 
	measured at the Tevatron, LHC, and an \nlc?  What is the discovery 
	reach for new strong dynamics, e.g., $t\bar{t}$ production through 
	the $\eta_{T}$ of technicolor or the $V_{8}$ of topcolor?

	\item  What are the consequences of the large Yukawa coupling of top 
	for the reactions $q\bar{q} \rightarrow (Z^{*},\gamma^{*}) 
	\rightarrow t\bar{t}H$ and $q\bar{q} \rightarrow Z^{*} \rightarrow 
	ZH, H\rightarrow t\bar{t}$ at the Tevatron, the LHC, and an \nlc\ 
	with c.m.\ energy of 0.5-1.5~TeV?
	
	\item What upper limits can experiment place on flavor-changing 
	neutral-current decays like $t \rightarrow c \gamma$?  How would 
	these limits constrain new physics?
	
	\item Does a muon collider have any special advantages for top 
	physics?

\end{enumerate}
\section*{Explanatory Notes}
\setlength{\parskip}{1.0ex}
More details appear in the transparencies for my talk at the Fermilab 
Workshop on Physics at a High-Luminosity Tevatron Collider, May 10, 
1996.  Here are comments and references for a few of the questions.
\frenchspacing

\noindent [3] E. Laenen, J. Smith, and W. van Neerven, 
\np{B369}{543}{92}, \pl{B321}{254}{94}; S. Catani, \etal, 
\hepph{9602208, 9604351}; E. L. Berger and H. Contopanagos, 
\hepph{9605212}.

\noindent [6] Advantageous bases for the analysis of spin 
correlations have been discussed by G. Mahlon and S. Parke, 
\pr{53}{4886}{96} and by S. Parke and Y. Shadmi (in preparation, 
available for Snowmass).  See the discussion of CP-nonconservation in 
D. Atwood, \etal,\hepph{9605345}.

\noindent [7] If, defying the expectations of the three-generation 
standard model, the lifetime of the top quark were $10\times$ greater 
than the canonical value of about $0.4 \times 10^{-24}\hbox{ s}$, top 
mesons would form.  The ground state (nearly degenerate pseudoscalar 
and vector states) would have a width around 150~MeV from the weak 
decay of the top.  Closely spaced $1^{+}$ and $2^{+}$ P-states 
lie 450~MeV above the ground state and decay by pion emission.  The 
width of the P-states will be 150~MeV from the weak decay of the top 
quark.  Might the nonobservation of the pion line in the $T\pi - T$ 
mass difference set a lower limit on the top width, hence an upper 
limit on the top lifetime?
\newpage
\noindent [9] Implications of the dead cone for the average 
	charged multiplicity in events containing heavy quarks are presented in 
	B. A. Schumm, Y. L. Dokshitzer, V. A. Khoze, and D. S. Koetke, 
	\prl{69}{3025}{92}. 
	For measurements of the multiplicity in tagged-$b$ 
	events on the $Z^0$ resonance, see R. Akers, \etal\ (OPAL 
	Collaboration), \zp{61}{209}{94}; K. Abe, \etal\ (SLD Collaboration), 
	\prl{72}{3145}{94}.  Tracks arising from $b$-decay are subtracted.

\noindent [10] For a recent survey, see G. W.-S. Hou, \hepph{9605203}.  Some 
of the prominent speculations are \textit{(i) a fourth generation + light 
scalars:} J.~F.~Gunion, D.~W.~McKay, and H. Pois, \pl{B334}{339}{94}, 
\pr{53}{1616}{96}; \textit{(ii) a fourth generation + supersymmetry:} M. 
Carena, H. E. Haber, and C. E. M. Wagner, \hepph{9512446}; \textit{(iii) an 
isocalar $Q=2/3$ quark:} V. Barger and R. J. N. Phillips, 
\pl{B335}{510}{94}; \textit{(iv) supersymmetry:} M. Carena, \etal, 
\np{B419}{213}{94}, \np{B426}{269}{94}; C. Kolda, \etal, 
\pr{50}{3498}{94}; V. Barger, \etal, \hepph{9404297}; J. Wells and G. 
Kane, \prl{76}{3498}{96}; S. Dimopoulos, \etal, \prl{76}{3494}{96}; 
S. Ambrosanio, \etal, \prl{76}{3498}{96} and \hepph{9605398}; 
\textit{(v) technipions:} 
K. Lane and E. Eichten, \pl{B222}{274}{89}; \textit{(vi) top-pions:} 
C. T. Hill, \pl{B266}{419}{91}, \ib{B345}{483}{95}.

\noindent [11] For the technicolor case, see E. Eichten and K. Lane, 
\pl{327}{129}{94}; for topcolor, see C. T. Hill and S. J. Parke, 
\pr{49}{4454}{94}.

\section*{Acknowledgment}
Fermilab is operated by Universities Research Association, Inc., under
contract DE-AC02-76CHO3000 with the United States Department of Energy. 
\end{document}